\documentclass[12pt]{elsart}
\usepackage{psfig,epsfig}
\usepackage[usenames]{color}
\journal{Nuclear Physics B}   %       LOCAL DEFINITIONS

\def\np#1#2#3{    {\it Nucl. Phys. }{\bf #1} (#2) #3}

\def\pl#1#2#3{    {\it Phys. Lett. }{\bf #1} (#2) #3}

\def\pr#1#2#3{    {\it Phys. Rev. }{\bf #1} (#2) #3}

\def\zp#1#2#3{    {\it Zeit. f\"ur Physik }{\bf #1} (#2) #3}

\def\re#1{{\mathrm Re}\left\{#1\right\} }
\def\im#1{{\mathrm Im}\left\{#1\right\} }
\newcommand{\com}[1]{ \par }

\def\evg{\, g_{eV}^\gamma}
\def\tvg{\, g_{\tau V}^\gamma}
\def\evz{\, g_{eV}^Z}
\def\eaz{\, g_{eA}^Z}
\def\tvz{\, g_{\tau V}^Z}
\def\taz{\, g_{\tau A}^Z}
\def\tz{\, g_{\tau T}^Z}
\def\tg{\, g_{\tau T}^\gamma}
\def\evg2{(g_{eV}^\gamma)^2}
\def\tvg2{(g_{\tau V}^\gamma)^2}
\def\evz2{(g_{eV}^Z)^2}
\def\eaz2{(g_{eA}^Z)^2}
\def\tvz2{(g_{\tau V}^Z)^2}
\def\taz2{(g_{\tau A}^Z)^2}
\def\tz2{(g_{\tau T}^Z)^2}
\def\tg2{(g_{\tau T}^\gamma)^2}

\def\re#1{{\mathrm Re}\left\{#1\right\} }
\def\im#1{{\mathrm Im}\left\{#1\right\} }

\def\dps{\displaystyle}
\newcommand{\beq}{\begin{equation}}
\newcommand{\eeq}{\end{equation}}
\newcommand{\bi}{\begin{itemize}}
\newcommand{\ei}{\end{itemize}}
\newcommand{\bea}{\begin{eqnarray}}
\newcommand{\eea}{\end{eqnarray}}
\newcommand{\bes}{\begin{eqnarray*}}
\newcommand{\ees}{\end{eqnarray*}}

\begin{document}
%\begin{tabular}{lr}
%\noindent
%FTUV$-$00$-$0218\hfill
%IFIC/00$-$17
%\end{tabular}
\begin{flushright}FTUV-06-1003\end{flushright}
\begin{frontmatter}
\title{CP violation and electric-dipole-moment
at low energy $\tau$ production with polarized electrons}
\author[Valencia]{J. Bernab\'eu},
\author[Montevideo]{G. A. Gonz\'alez-Sprinberg} and
\author[Valencia]{J. Vidal}
\address[Valencia]{Departament de F\'{\i}sica Te\`orica
Universitat de Val\`encia, E-46100 Burjassot,Val\`encia, Spain\\
and\\%}
%\address[ific]{
IFIC, Centre Mixt Universitat de Val\`encia-CSIC, Val\`encia, Spain}
\address[Montevideo]{Instituto de F\'{\i}sica,
 Facultad de Ciencias, Universidad de la Rep\'ublica,
 Igu\'a 4225, 11400 Montevideo, Uruguay}
 %\journal{the 2nd Workshop on Super B-Factory}
 \begin{abstract}
The new proposals for high luminosity B/Flavor factories, near and
on top of the $\Upsilon$ resonances,
 allow for a detailed investigation of $CP$-violation
in the $\tau$-pair production. In particular, bounds on the tau
electric dipole moment can be obtained from genuine CP-odd
observables related to the $\tau$-pair production. We perform an
 independent analysis from low energy (10 GeV) data
 %from the high energy data 
 by means of
 %low energy 
linear spin observables.  We show that for a
longitudinally polarized electron beam a  $CP$-odd asymmetry,
associated to the normal polarization term, can be measured at
these low energy facilities both at resonant and non resonant
energies. In this way stringent and independent bounds to the tau
electric dipole moment, which are orders of magnitude below other
 high or low energy bounds, can be
 obtained.
\end{abstract}
\end{frontmatter}

\section{Introduction}
The standard model describes with high accuracy most of the
physics found in present experiments \cite{pdg}. Nowadays,
however, neutrino physics offers a first clue to physics beyond
this "low energy" model \cite{nu}. Other signals of new phenomena
may also appear in CP violation physics. While CP violation in the
standard model can be easily introduced by  quark mixing, as in
the CKM mechanism, the discovery of CP violation in the lepton
sector would establish new sources of CP violation and the
appearance of new physics. The time reversal odd electric dipole
moment (EDM) of the $\tau$ is the source of CP violation in the
$\tau$-pair production vertex. In the framework of local quantum
field theories the CPT theorem states that CP violation is
equivalent to T violation. While the T-odd electric dipole moments
(EDM) of the electron and muon have been extensively investigated
both in experiment and theory,
 the case of the tau is somewhat different. The dipole moment effective
 operators flip
 chirality and are therefore related to the mass mechanism of
  the theory. The tau
lepton has a relatively high mass: this means that tau lepton
physics is expected   to be more sensitive to contributions to
chirality-flip terms coming from high energy scales and new
physics. Furthermore, the tau has a very short lifetime
and can decay  into hadrons, so different techniques to those for
the (stable) electron or muon case are needed in order to measure
the dipole moments.
 There are very precise  bounds on the EDM magnitude
 of nucleons and leptons;  the most precise one is the electron
EDM, $d^e_\gamma = (0.07\pm0.07) \times 10^{-26}$ {\it e cm},
 while the looser one is the  $\tau$ EDM \cite{pdg},
$-0.22 \, \it{ e \, cm } < Re (d^\tau_\gamma) \times 10^{16} <
0.45\, \it{e\, cm}$.  From the theoretical point of view the CP
violation mechanisms in many models provide a kind of accidental
protection in order to generate an EDM for quarks and leptons.
This is the case in the CKM mechanism, where EDM and weak-EDM are
generated only at very high order in the coupling constant. This
opens a way to  test many models: $CP$-odd observables related to
EDM would give no appreciable effect from the standard model and
any experimental signal should be identified with  beyond the
standard model physics. Following the ideas of \cite{nt} and
\cite{heidel}, the tau weak-EDM has been studied in $CP$-odd
observables \cite{l3,ao} at high energies through linear
polarizations and spin-spin correlations. EDM bounds for the tau,
from $CP$-even observables such as total cross sections or decay
widths, have also been considered in \cite{paco,grif,masso}. In
ref.\cite{rindani} the sensitivity to the WEDM in spin-spin
correlation observables was studied for tau-charm-factories with
polarized electrons. While most of the statistics for the tau pair
production was dominated in the past by high energy physics,
mainly at LEP, nowadays the situation has evolved. High luminosity
B factories and their upgrades  at resonant energies ($\Upsilon$
thresholds) have the largest $\tau$ pair samples. In the future,
the possibility for a Super B/Flavor Factory with a $\tau$-pair production 
rates many orders of magnitude higher than present samples is being intensively
analyzed \cite{super}. These facilities may also have the
possibility of polarized beams. This calls for a dedicated study
of the observables related to CP violation
 and the EDM of the $\tau$ lepton at low energies.
In this paper we study a set of different observables for the tau
system that may lead to competitive results  with the present
bounds for the EDM.

\section{Effective Lagrangian}

Deviations from the standard model, at low energies, can be
parametrized by an effective Lagrangian built with the standard
model particle spectrum, having as zero order term just the
standard model Lagrangian, and containing higher dimension gauge
invariant operators suppressed by the scale  of new physics,
$\Lambda$ \cite{buch}. The leading non-standard effects come from
the operators with the lowest dimension. For CP violation those are
dimension six operators and there are only two operators of this
type that contribute \cite{arcadi} to the tau EDM and weak-EDM:
\begin{eqnarray}
\label{eq:ob}
\mathcal{O}_B = \frac{g'}{2\Lambda^2} \overline{L_L} \varphi \sigma_{\mu\nu}
 \tau_R B^{\mu\nu} ~,& \hspace*{1cm} \mathcal{ O}_W
 = \frac{g}{2\Lambda^2} \overline{L_L} \vec{\tau}\varphi
\sigma_{\mu\nu} \tau_R \vec{W}^{\mu\nu}  ~.
\end{eqnarray}
Here $L_L=(\nu_L,\tau_L)$ is the tau leptonic doublet,
$\varphi$  is the Higgs doublet, $B^{\nu\nu}$ and $\vec{W}^{\mu\nu}$ are the
U(1)$_Y$ and SU(2)$_L$ field strength tensors, and $g'$ and $g$ are the gauge
couplings.

Other possible operators that one could imagine
reduce to the above ones of Eq.(\ref{eq:ob})  after
using the standard model equations of motion.
In so doing, the couplings will be proportional
to the tau-lepton Yukawa couplings.

Thus, the effective Lagrangian for the EDM is:
\begin{equation}
\label{eq:leff}
\mathcal{ L}_{eff} = i \alpha_B \mathcal{ O}_B + i
\alpha_W \mathcal{ O}_W + \mathrm{h.c.}
\label{eq:interaccio}
\end{equation}
where the couplings
$\alpha_B$ and $\alpha_W$ are real. Note that complex couplings do not break
$CP$ conservation and lead to magnetic dipole moments which are not considered
in this paper where we are mainly interested on $CP$-odd observables.

After spontaneous symmetry breaking the neutral scalar gets a
vacuum expectation value and the interactions in Eq.(\ref{eq:interaccio}) 
can be written in
terms of the gauge boson mass eigenstates $A^\mu$ and $Z^\mu$ as:

\begin{eqnarray}
\mathcal{ L}_{eff}^{\gamma, Z} &=& - i  \,d^\tau_\gamma
\,\overline{\tau} \sigma_{\mu\nu}
 \gamma^5 \tau F^{\mu\nu} -
 i  \,d^\tau_Z \,\overline{\tau} \sigma_{\mu\nu}
\gamma^5 \tau
Z^{\mu\nu}%\\
\label{eq:leff_fin}
\end{eqnarray}
where $F_{\mu\nu}=\partial_\mu A_\nu-\partial_\nu A_\mu$
  and
$Z_{\mu\nu}=\partial_\mu Z_\nu-\partial_\nu Z_\mu$
are  the abelian field strength tensors of the photon and
  $Z$ gauge boson and $d^\tau_\gamma$ and $\d^\tau_Z$ are the electric and
  weak-electric dipole moments, respectively.  We have not written 
  in Eq.(\ref{eq:leff_fin}) some of the terms coming from
Eq.(\ref{eq:interaccio}) because they do not contribute at leading
order to the observables we are interested in. These terms are the
non-abelian couplings involving more than one gauge boson and the
term related to the $CP$-odd $\nu_\tau - \tau - W^\pm$ couplings.
In the effective Lagrangian approach the same couplings that
contribute to
 the EDM  form factor,
$F^{\mathrm new}(q^2)$, also contribute to the EDM defined at
$q^2=0$. Only higher dimension operators contribute to the
difference $F^{\mathrm new}(q^2)-F^{\mathrm new}(0)$ and, if $
|q^2| \ll \Lambda^2$, as required for the consistence of the
effective Lagrangian approach, their effects will be suppressed by
powers of $q^2/\Lambda^2$. This allows us to make no distinction
between the electric {\it dipole moment} and the electric {\it
form factor} in this paper.

\begin{figure}[hbtp]
\begin{center}
\epsfig{file=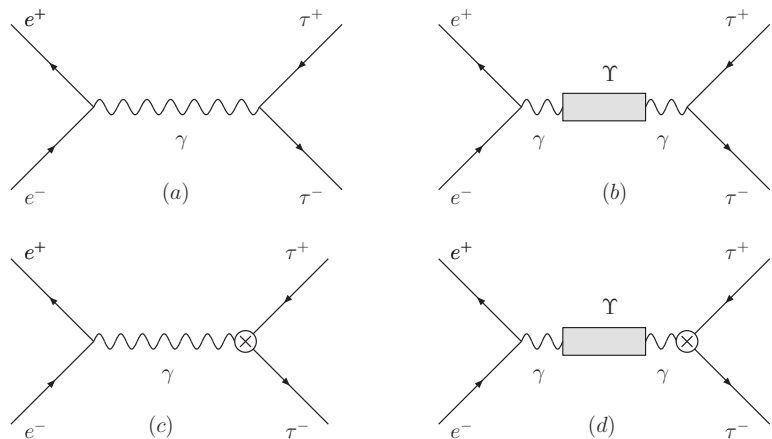,width=0.75\textwidth}
\end{center}
\caption{Diagrams (a) direct $\gamma$ exchange (b) $\Upsilon$
production (c) EDM in $\gamma$ exchange (d) EDM at the
$\Upsilon$-peak.} \label{fig:figura1}
\end{figure}

The $e^+\, e^- \longrightarrow
\tau^+ \tau^-$ cross section has contributions coming from
the standard model and the effective Lagrangian
Eq.(\ref{eq:leff_fin}). At low energies the
tree level contributions come from
  $\gamma$ exchange  
or $\Upsilon$ 
exchange in the s-channel.
The interference with the $Z$-exchange
($\gamma - Z$,  $\Upsilon - Z$ at the $\Upsilon$ peak) and the
$Z-Z$ diagrams are suppressed by powers of $\left(q^2/M_Z^2\right)$. The tree level
contributing diagrams are shown in  Fig.\ref{fig:figura1} where diagrams ($a$) and
($b$) are standard model contributions, and ($c$)and ($d$) come from beyond the standard
model terms in the Lagrangian. Notice that
standard model radiative corrections
that may contribute to $CP$-odd observables (for example, the ones
                                                 that generate the
standard model electric dipole moment for the $\tau$) come in
higher order in the  coupling constant, and at present level of
experimental sensitivity they are not measurable. On these grounds
the  bounds on the EDM that one may get are just the ones coming
from  beyond the standard model physics. Taking into account that our
observables are genuine $CP$-odd, higher order $CP$-even amplitudes can be 
neglected.
%%%%%%%%%%%%%
%%%%%%%%%%%%%%%

\section{Low energy polarized beams and the EDM.}

As we will show, the EDM  can be studied  at leading order in the angular
distribution of the $e^+e^-  \longrightarrow
\tau^+(s_+)\tau^-(s_-)$ differential cross section for
longitudinally polarized electrons. The polarization of the final
fermions is determined through the study of the angular
distribution of their decay products. In our analysis we only keep
linear terms in the EDM, neglecting terms proportional to the mass
of the electron.

When considering the measurement of the polarization of just one of the taus,
the normal -to the scattering plane- polarization ($P_N$) of each tau
 is the only
component which is $T$-odd. For $CP$-conserving interactions, the
$CP$-even term $(s_++s_-)_N$ of the normal polarization only gets
contribution through the combined
effect of both an helicity-flip transition and the presence of
absorptive parts, which are both
suppressed in the standard model. For a $CP$-violating interaction,
such as an EDM, the $(s_+-s_-)_N$ $CP$-odd term  gets a
non-vanishing value without the need of absorptive parts.

As $P_N$ is even under parity ($P$) symmetry, the observable
sensitive to the EDM will need, in addition to $d_\gamma^\tau$,
an additional $P$-odd contribution coming  from  longitudinally polarized
electrons. A standard axial coupling, coming from a
$Z$-exchange in the s-channel, could also be considered as an alternative 
but in that case the
contribution is suppressed by powers of
$\left(q^2/M_Z^2\right)$.

Following the notation of references \cite{arcadi} and \cite{nos}, we
now show how to  measure the EDM using low energy $CP$-odd observables.
 Our aim is to identify genuine $CP$-odd
 observables that are linear in the EDM and
not (additionally) suppressed by either $(q^2/M_Z^2)$ or unitarity
corrections.

In the center of mass (CM) reference frame we choose the
coordinates as in Fig.(2).
The $\mbox{\boldmath $s$}^\pm$  are the $\tau^\pm$  spin vectors in
the $\tau^\pm$ rest system, $s_\pm=(0, s^x_\pm, s^y_\pm, s^z_\pm)$.
With this setting, polarization
along the directions
$x,y,z$ correspond to what is called transverse
 (T), normal (N) and longitudinal (L)
polarizations, respectively.

We  first consider the $\tau$-pair production in $e^+e^-$
collisions though direct $\gamma$ exchange (diagrams (a) and (c)
in Fig. \ref{fig:figura1}.). Next,  we will show that the basic
results of this section still hold for resonant $\Upsilon$
production.

Let us  assume from now on that the tau production plane and
direction of flight can be fully reconstructed. This can be  done
\cite{kuhn} if both $\tau$'s decay semileptonically.
The differential cross section for $\tau$ pair production with
polarized electrons with helicity $\lambda$  is:

\begin{equation}
\left. \frac{d \sigma}{d  \Omega_{\tau^-}}\right|_\lambda=
\left.\frac{d \sigma^{0}}{d  \Omega_{\tau^-}}\right|_\lambda
+\left. \frac{d \sigma^{S}}{d  \Omega_{\tau^-}}\right|_\lambda+\ldots
\label{cross1}
\end{equation}

The terms to be considered include contributions from leading order standard 
model and effective operator (EDM).
The dots take account for  higher order terms in
the effective Lagrangian that are beyond experimental sensitivity and which are not
considered in this paper.

The first term of Eq. (\ref{cross1}) represents the $\tau$ spin-independent
differential cross section. The second term $\displaystyle
\left.\frac{d\sigma^{S}}{d\Omega_{\tau^-}}\right|_\lambda$ includes the linear
terms in the spin of the $\tau$'s and has sensitivity to the EDM in their normal
polarization:

\bea
\left.\frac{d \sigma^{S}}{d
\Omega_{\tau^-}}\right|_\lambda=\frac{\alpha^2}{16\ s}\, \beta &&
\left\{\lambda\left[
(s_-+s_+)_xX_++(s_-+s_+)_zZ_++(s_--s_+)_yY_- \right)]+\right.\nonumber \\
&&\quad \left. (s_--s_+)_x X_- + (s_--s_+)_z Z_-\right\}
\eea
where
\bea
X_+&=&\frac{1}{\gamma}\sin\theta_{\tau^-},\qquad\qquad\qquad\qquad 
X_-=-\frac{1}{2}\sin(2\theta)\,
\dps\frac{2m_\tau}{e}\im{d_\tau^\gamma},\nonumber\\
Z_+&=&-\cos \theta_{\tau^-},\qquad\qquad\qquad\qquad 
Z_-=-\frac{1}{\gamma}\sin^2\theta\,
\dps\frac{2m_\tau}{e}\im{d_\tau^\gamma},\nonumber\\
Y_-&=&\gamma\beta^2\cos \theta_{\tau^-}\, \sin \theta_{\tau^-}\,
\dps\frac{2m_\tau}{e}\re{d_\tau^\gamma}. \label{cross3}
\eea
and $\alpha$ is the fine structure constant, $s=q^2$ is the squared CM energy 
and $\gamma=\frac{\sqrt{s}}{2
m_\tau}$, 
$\beta=\sqrt{1-\frac{1}{\gamma^2}}$ are the dilation factor and
$\tau$ velocity, respectively.

As can be seen in Eq.(\ref{cross3}), the EDM is the leading
contribution to the Normal Polarization of the single tau.

%%%%%%%%%%%%%%%%%%%%%%%%%%%%
\begin{figure}[hbtp]
\begin{center}
\epsfig{file=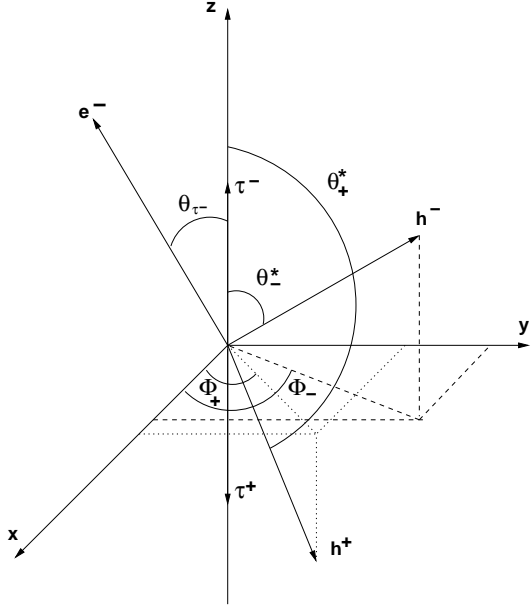,width=0.5\textwidth}
\end{center}
\caption{Coordinate system for $h^\pm$ production from the $\tau^\pm$}
\label{fig:fig1}
\end{figure}
%%%%%%%%%%%%%%%

\subsection{Single $tau$ normal polarization observable}

We now show how to get an observable proportional to the EDM term
from the Normal polarization of a single tau in the complete process

\[ e^+e^-(pol) \rightarrow \gamma \rightarrow \tau^+\tau^-
\rightarrow h^+\bar{\nu}_\tau h^-\nu_\tau
\]
The cross section can be written as a function of the kinematical 
variables of the
hadrons into which each tau decays \cite{tsai} as:
\begin{eqnarray}
&&\hspace*{-1cm}d\sigma \left.\left(e^+e^-\rightarrow \gamma
\rightarrow \tau^+\tau^- \rightarrow h^+\bar{\nu}_\tau h^-\nu_\tau\right)\right|_\lambda=
4\, d\sigma
\left.\left(e^+e^- \rightarrow \tau^+(\overrightarrow{n}_+^*)
\, \tau^-(\overrightarrow{n}_-^*)\right)\right|_\lambda\nonumber \\
&&\times \, Br(\tau^+ \rightarrow h^+\bar{\nu}_\tau)
Br(\tau^- \rightarrow h^-\nu_\tau)
\frac{d\Omega_{h^+}}{4\pi}\, \frac{d\Omega_{h^-}}{4\pi}
\label{eq:cros1}\end{eqnarray}
with 
\beq \overrightarrow{n}_\pm^*= \mp\alpha_\pm
\frac{\overrightarrow{q}^{  *}_\pm}{
\arrowvert\overrightarrow{q}^{  *}_\pm\arrowvert} =
\mp\alpha_\pm(\sin\theta_{\pm}^*\, \cos\phi_\pm,
\sin\theta_{\pm}^*\, \sin\phi_\pm,\cos\theta_{\pm}^*)\nonumber\\
\eeq

The linear terms in the spin of the taus
depend on  several kinematic variables that we have to take
into account: the CM polar angle $\theta_{\tau^-} $ of production
of the $\tau^-$ with respect to the electron, the azimuthal
$\phi_{\pm}$ and polar $\theta^*_{\pm}$ angles of the 
produced hadrons $h^\pm$ ($\hat{q}^*_{\pm}$) in the $\tau^\pm$
rest frame (the * means that the quantity is
 given in the $\tau$ rest frame; see Fig.\ref{fig:fig1}).
These angles appear in a different way on each term. The
$\theta_{\tau^-}$ angle enters in  the cross section as
coefficients (of the  $(s_-+s_+)_x$ term, for example) while  the
hadron's angles appear in the cross section through the polarization parameters
$\overrightarrow{n}^*_\pm$.
The whole angular dependence of each contribution is unique and it
is this dependence that allows
 to select one of the terms in the cross section.
Indeed, it is by an integration on the $\theta_{\tau^-}$ angle, followed by a
dedicated integration on the hadronic angles, that
one can select a polarization term and there, the contribution of the EDM.

For the Normal Polarization term this works as follows.
The integration over the $\tau^-$ variables $d\Omega_{\tau^-}$ erases all
the information on the $Z_+$ and $Y_+$ term of
the cross section. Then, the cross
section can be written only in terms of
the surviving terms as:

\begin{eqnarray}
\left.d^4\sigma^{S}\right|_\lambda &=&
\frac{\pi^2\alpha^2\beta}{4\, s}
\, Br(\tau^+ \rightarrow h^+\bar{\nu}_\tau)
Br(\tau^- \rightarrow h^-\nu_\tau) \, \frac{d\Omega_{h^+}}{4\pi}\,
\frac{d\Omega_{h^-}}{4\pi}\, \times \nonumber\\
&&\Bigg\{\frac{\lambda}{\gamma}\, \left[(n_-^*)_x+(n_+^*)_x\right] +\lambda\,
\gamma\beta\, \left[(n_-^*)_y-(n_+^*)_y\right]\frac{2 m_\tau}{e}\,
\re{d^\gamma_\tau}\\
&&+\frac{4}{3\gamma}\left[(n_-^*)_z-(n_+^*)_z\right]\frac{2 m_\tau}{e}\,
\im{d^\gamma_\tau}\Bigg\} \label{eqxy1}
\end{eqnarray}

As can be seen up to this point, helicity independent normal (longitudinal) polarization
terms due to absorptive (EDM imaginary) parts  may also
survive. One may get rid of them by subtracting the cross sections for
different helicities.
\beq
\left.\d^2\sigma^{S}\right|_{Pol( e^-)}\equiv \left.d^4\sigma^{S}\right|_{\lambda=1}-
\left.d^4\sigma^{S}\right|_{\lambda=-1}\label{spol1}
\eeq
Then,
in order to enhance and select the corresponding $P_N$ observable, one has
integrate as much as kinematic variables as possible without erasing the
signal of the EDM ($\re{d^\gamma_\tau}\rightarrow d^\gamma_\tau$ from now on).
Keeping only azimuthal angles and integrating all
other variables one gets:

\bea
\left.\frac{d^2\sigma^{S}}{d\phi_- d\phi_+}\right|_{Pol( e^-)}
&=&   \frac{\pi\alpha^2\beta}{32 s}\,
Br(\tau^+ \rightarrow h^+\bar{\nu}_\tau)
Br(\tau^- \rightarrow h^-\nu_\tau)
  \times  \nonumber\\
 & & \Big\{\frac{1}{\gamma}\left[(\alpha_-) \cos\phi_--(\alpha_+)\cos\phi_+\right]+
 \label{eqxy3} \\
&&\beta\, \gamma  \left[(\alpha_-) \cos\phi_--(\alpha_+)\cos\phi_+\right]\frac{2
m_\tau}{e}\,d^\gamma_\tau  \Big\}
\eea

Now, to be sensitive only to the EDM  we can define the
azimuthal asymmetry as:

\begin{equation}
A_N^{\mp} =\frac{\sigma^\mp_{L} - \sigma^\mp_{R}}{\sigma}=
\alpha_\mp\frac{3\pi\gamma\beta}{8(3-\beta^2)}\frac{2
m_\tau}{e}\,d^\gamma_\tau\label{asym}
\end{equation}

where

\begin{eqnarray}
\sigma^\mp_L &=& \int_0^{2\pi}d\phi_\pm \left[\int_0^\pi d\phi_\mp\,\left.
\frac{d^2\sigma^S}{d\phi_- d\phi_+}\right|_{Pol (e^-)}\right]=\nonumber\\
&&\quad Br(\tau^+ \rightarrow h^+\bar{\nu}_\tau)
Br(\tau^- \rightarrow h^-\nu_\tau)\,\alpha_\mp\frac{(\pi\alpha\beta)^2\gamma}{8s}\,
\frac{2m_\tau}{e}\,d^\gamma_\tau\\
\sigma^\mp_R &=& \int_0^{2\pi}d\phi_\pm \left[\int_\pi^{2\pi}d\phi_\mp\,\left.
\frac{d^2\sigma^S}{d\phi_- d\phi_+}\right|_{Pol (e^-)}\right]=\nonumber \\
&&\quad -Br(\tau^+ \rightarrow h^+\bar{\nu}_\tau)
Br(\tau^- \rightarrow h^-\nu_\tau)\,\alpha_\mp\frac{(\pi\alpha\beta)^2\gamma}{8s}\,
\frac{2m_\tau}{e}\,d^\gamma_\tau\
\end{eqnarray}

It is easy to verify that all other terms in the considered cross
section are eliminated when we integrate in this way.
Notice that this integration procedure does not erase
contributions coming from the $CP$-even term of the Normal
Polarization as it will be shown in the next section.

To get rid of $CP$-even terms one has to define a true $CP$ violation observable
by summing up the defined
asymmetry (\ref{asym}) for $\tau^+$ and for $\tau^-$
\beq
A_N^{CP}=\frac{1}{2}\left(A_N^++A_N^-\right)= \alpha_h
\frac{3\pi\gamma\beta}{8(3-\beta^2)}\frac{2
m_\tau}{e}\,d^\gamma_\tau\label{asimcp}
\eeq

This observable is free of the CP-even contributions  described in
what follows and it is a genuine CP-odd observable.

\subsection{$\gamma-Z$ interference}

Contributions to this observable can also come from
the standard $Z-\gamma$ interference:

\bea
\left.\frac{d \sigma^0}{d  \Omega_{\tau^-}}\right|_\lambda^{Z\gamma}
&=&\frac{\alpha^2 |P_Z|^2}{16\ (2s_wc_w)^2}\, \beta\,
\left(s-M_Z^2\right)\, M^{Z\gamma}_0\\
\left.\frac{d \sigma^S}{d  \Omega_{\tau^-}}\right|_\lambda^{Z\gamma}
&=&\frac{\alpha^2 |P_Z|^2}{16\ (2s_wc_w)^2}\, \beta
\left\{\Gamma_Z\, M_Z \left[(s_-+s_+)_y Y_+^{Z\gamma}\right]+\right.\nonumber\\
&&\left.\left(s-M_Z^2\right)\left[(s_-+s_+)_xX^{Z\gamma}_+
+(s_-+s_+)_zZ^{Z\gamma}_+\right]+\right\} \eea where \bea |P_Z|^2
&=&\frac{1}{(s-M_Z^2)^2+\Gamma_Z^2M_Z^2},\quad
a=-\frac{1}{2},\quad
v=-\frac{1}{2}+2s_w^2\nonumber \\
M_0^{Z\gamma}&=&\lambda\, a\,
v(2\beta\cos\theta+2-\beta^2\sin^\theta)-a^2\beta\cos\theta-v^2(2-\beta^2\sin^2\theta)\nonumber
\\
X_+^{Z\gamma}&=&\frac{1}{\gamma}\left[\lambda(2v^2+a^2\beta\cos\theta)-a\,
v(2-\beta\cos\theta)\right]\nonumber \\
Z^{Z\gamma}_+&=&\left[a^2\beta(1+\cos^2\theta)+2v^2\cos\theta\right]-a\, v
\left[\beta(1+\cos^2\theta)+2\cos\theta\right]\nonumber \\
Y_+^{Z\gamma}&=& \frac{1}{\gamma}a\,\beta(\lambda\, v-a)\sin\theta
\label{crosz}
\eea

Subtracting the cross sections for different helicities and
integrating, as in the previous section, one gets \bea
^{Z\gamma}\sigma^\mp_L &=& \frac{\pi \alpha^2\beta}{2(2s_wc_w)^2
s}\, |P_Z|^2\, Br(\tau^+ \rightarrow h^+\bar{\nu}_\tau) Br(\tau^-
\rightarrow h^-\nu_\tau)\, a\, v\, \times
\nonumber \\
&&\quad\left\{ 4 \left(s-M_Z^2\right)\left(1-\frac{\beta^2}{3}\right)\mp \alpha_\mp\Gamma_Z M_Z
\frac{\pi\beta}{4\gamma}\right\}\\
^{Z\gamma}\sigma^\mp_R &=& \frac{\pi \alpha^2\beta}{2(2s_wc_w)^2 s}\, |P_Z|^2\,
Br(\tau^+ \rightarrow h^+\bar{\nu}_\tau) Br(\tau^- \rightarrow h^-\nu_\tau)\,
a\, v\, \times
\nonumber \\
&&\quad\left\{ 4 \left(s-M_Z^2\right)\left(1-\frac{\beta^2}{3}\right)\pm \alpha_\mp\Gamma_Z M_Z
\frac{\pi\beta}{4\gamma}\right\}
\eea
so that the $Z$ contribution to the asymmetry (\ref{asym}) is

\beq
^{Z\gamma}A_N^{\mp} =\mp \alpha_\mp\frac{3\pi\beta\gamma}{8(3-\beta^2)}\,
\overbrace{\frac{a\, v}{\gamma^2(2s_wc_w)^2}
\frac{s\Gamma_Z M_Z}{(s-M_Z^2)^2+(\Gamma_Z M_Z)^2}}^\varepsilon
\label{asym1}
\eeq

 At 10 Gev, the value of the $\varepsilon$ factor is of the order
 $1\times10^{-6}$, which makes this contribution to the asymmetry two
 orders of magnitude below the expected sensitivity for the EDM. Anyway, this
 asymmetry does not contribute to the $CP$-odd $A_N^{CP}$ of Eq.(\ref{asimcp}).

\subsection{Observables at the $\Upsilon$ resonances}

All these ideas can be applied for $e^+e^-$ collisions at the $\Upsilon$ peak
where the $\tau$ pair production is  mediated by the resonance:
$e^+e^- \rightarrow \Upsilon \rightarrow \tau^-\tau^-$.
At the $\Upsilon$ production energies  we
 have an important  tau pair production rate.
We are interested in $\tau$ pairs produced by the
 decays of the $\Upsilon$ resonances, therefore
we can use $\Upsilon(1S)$, $\Upsilon(2S)$ and $\Upsilon(3S)$ where
the decay rates into tau pairs have been measured. At the
$\Upsilon(4S)$ peak, although  it decays dominantly into
$B\overline{B}$, high luminosity B-Factories have an important
direct tau pair production. Except for this last case, that can be
studied with the results of the preceding sections,  we assume
that the resonant diagrams (b) and (d) of Fig. 1. dominate the
process on the $\Upsilon $ peaks. This has been extensively
discussed in ref.(\cite{Bernabeu:2004ww}.

The main result is that the tau pair production at the $\Upsilon$
peak introduces the same tau polarization matrix terms as the
direct production with $\gamma$ exchange (diagrams (a) and (c)).
The only difference is an overall factor $|H(s)|^2$  in the cross
section which is responsible for the enhancement at the resonant
energies, the pure resonant (imaginary) amplitude being
\beq H(M_\Upsilon^2) = -i \, \frac{3}{\alpha}
Br\left(\Upsilon \rightarrow e^+e^-\right) \label{factor} 
\eeq

Besides, it is easy
to show that, at the $Upsilon$ peak,  the interference of diagrams
(a) and (d) plus the interference of diagrams (b) and (c) is
exactly zero and so it is the interference of diagrams (a) and
(b). Finally, the only contributions proportional to the   EDM
come  with the interference of diagrams (b) and (d), while diagram
(b) squared gives the leading contribution to the cross section.

The computations we did before can be repeated here,
and finally we obtain no changes in the asymmetries:
 the only difference  is in the value of the resonant production cross section
at the $\Upsilon$ peak that is multiplied by the overall factor
$|H(M_\Upsilon^2)|^2$.

In fact,  one can take the four diagrams (a,b,c,d) together and
still get the same results we have already shown.
Energies {\it off} or {\it on} the resonance will
automatically select the significant diagrams.

\section{Bounds on the EDM }
We can now estimate the bounds on the EDM that can be achieved
using this observable. For numerical results we assume a set of
integrated luminosities  for high statistics $B$/Flavor factories.
We also consider the $\pi^\pm \; \bar{\nu_\tau}$ or $\rho^\pm\;
\bar{\nu_\tau}$ ({\it i.e.} $h_1 , \;h_2= \pi ,\; \rho$ ) decay
channels for the traced $\tau^\pm$,
while we sum up over $\pi^\mp \; \nu_\tau$ and $\rho^\mp\;
  \nu_\tau$ hadronic decay channels
for the non traced $\tau^\mp$.

For comparison with other references, we show the  bounds for the
$\tau$-EDM that can be set in different scenarios:
\begin{eqnarray}
\hspace*{-.3cm}|d^\gamma_\tau|&\le& 4.4\ 10^{-19}\ e cm  \quad \mbox{Babar + Belle at $2 ab^{-1}$}\nonumber\\
\hspace*{-.3cm}|d^\gamma_\tau|&\le& 1.6\ 10^{-19}\ e cm \quad
\mbox{Super
B/Flavor  factory, 1 yr running, $15 ab^{-1}$}\nonumber\\
\hspace*{-.3cm}|d^\gamma_\tau|&\le& 7.2\ 10^{-20}\ e cm  \quad
\mbox{Super B/Flavor factory, 5 yrs running, $75 ab^{-1}$}
\end{eqnarray}

These bounds improve present ones by 3 orders of
magnitude.

We can also define other observables that are sensitive to the
imaginary part of the EDM. The analysis is similar to the one we
have done here and in ref.(\cite{Bernabeu:2004ww}).

To conclude, we have shown that low energy data makes possible a clear 
separation of the
effects coming from the electromagnetic-EDM, the weak-EDM and
interference effects. Polarized electron beams open the possibility to put
bounds on the $\tau$ EDM looking at single tau polarization
observables with low energy data. These observables allow for an
independent analysis of the EDM bounds from what has been done
with other high and low energy data.

\begin{ack}
This work has been supported by CONICYT-PDT-094-Uruguay, by MEC
and FEDER, under the grants FPA2005-00711 and FPA2005-01678, and
by Gene\-ralitat Valenciana under the grant GV05/264.
\end{ack}
%%%%%%%%%%%%%%%%%%%%%%%%%%%%%%%%%%%%%%%%%


\begin{thebibliography}{20}

\bibitem{pdg}  W.-M. Yao et al., Journal of Physics G {\bf 33}, (2006) 1;
 K.~Inami {\it et al.} [BELLE Collaboration],
\pl{B551}{2003}{16}.

\bibitem{nu}
Y.~Fukuda {\it et al.} [Super-Kamiokande Collaboration], Phys.\
Rev.\ Lett.\ {\bf 81} (1998) 1562.

\bibitem{nt} J.~Bernab\'eu, G.A.~Gonz\'alez-Sprinberg and J.~Vidal,
\pl{B326}{1994}{168}.

\bibitem{heidel} W.~Bernreuther, U.~Low, J.~P.~Ma and O.~Nachtmann,
Z.\ Phys.\ C {\bf 43}, 117 (1989).

\bibitem{l3} M. Acciarri {\it et al.} [L3 Collaboration], \pl{B434}{1998}{169}.

\bibitem{ao} D.~Buskulic {\it et al.}  [ALEPH Collaboration],
Phys.\ Lett.\ B {\bf 346}, 371 (1995);
K.~Ackerstaff {\it et al.}  [OPAL Collaboration],
Z.\ Phys.\ C {\bf 74}, 403 (1997); H.Albrecht {\it et al.} [ARGUS Collaboration], \pl{B485}{2000}{37}.

\bibitem{paco} F. del Aguila and M. Sher, \pl{B252}{1990}{116}.

\bibitem{grif} J.A. Grifols and A. Mendez, \pl{B255}{1991}{611} and Erratum
\pl{B259}{1991}{512}.

\bibitem{masso} R. Escribano and E. Masso, \pl{395}{1997}{369}.

\bibitem{rindani} B. Ananthanarayan and S.D. Rindani, \pr{D51}{1995}{5996}

\bibitem{super} See for example
http://www.lnf.infn.it/conference/superb06/ and
http://www-conf.slac.stanford.edu/superb/.

\bibitem{buch}W.~Buchmuller and D.~Wyler, \np{B268}{1986}{621};
C.N.~Leung, S.T.~Love and S.~Rao, \zp{C31}{1986}{433}; %
M. Bilenky and A. Santamaria, \np{B420}{1994}{47}.

\bibitem{arcadi}  G.A.~Gonz\'alez-Sprinberg, A.~Santamaria, J.~Vidal, \np{B582}{2000}{3}.

%\bibitem{500}
%W.~Bernreuther, O.~Nachtmann and P.~Overmann,
%Phys.\ Rev.\ D {\bf 48} (1993) 78.

%\bibitem{nuria}J. Bernab\'eu and N. Rius,\pl{B232}{1989}{127};
%J. Bernab\'eu, N. Rius and A. Pich,\pl{B257}{1991}{219}.

\bibitem{nos} J.~Bernab\'eu, G.A.~Gonz\'alez-Sprinberg, M.~Tung and J.~Vidal,
\np{B436}{1995}{474}.

\bibitem{kuhn}
J.~H.~Kuhn, Phys.\ Lett.\ B {\bf 313} (1993) 458.


\bibitem{tsai} Y.S.~Tsai \pr{D4}{1971}{2821}.

\bibitem{Bernabeu:2004ww}
  J.~Bernabeu, G.~A.~Gonzalez-Sprinberg and J.~Vidal,

  Nucl.\ Phys.\ B {\bf 701}(2004) 87.

\end{thebibliography}
\end{document}